\documentclass[a4paper,11pt]{scrartcl}

\usepackage[utf8]{inputenc}
\usepackage{hyperref}
\usepackage{amsmath}
\usepackage{amssymb}
\newtheorem{theorem}{Theorem}

\usepackage{graphicx}
\graphicspath{{figs/}}

\usepackage[font=small,labelfont=bf]{caption}
\usepackage[font=small,labelfont=normalfont,labelformat=simple]{subcaption}

\usepackage[usenames,dvipsnames]{xcolor}
\newcommand{\remrkblue}[2]{\textcolor{blue}{\textsc{#1:}}\textcolor{blue}{\textsf{#2}}}
\newcommand{\manfred}[2][]{\remrkblue{Manfred #1}{#2}}

\DeclareMathOperator{\sgn}{sgn}

\title{Many Order Types on  Integer Grids of~Polynomial Size\footnote{
		The author acknowledges support by the internal research funding ``Post-Doc-Funding'' from Technische Universit\"at Berlin.}}
\author{Manfred Scheucher}
\date{}

\def\inst#1{$^{#1}$}

\begin{document}

\author{
Manfred Scheucher\inst{12}
}

\maketitle

\vspace{-1cm}

\begin{center}
{\footnotesize
\inst{1} 
Institut f\"ur Mathematik, \\
Technische Universit\"at Berlin, Germany,\\
\texttt{\{scheucher\}@math.tu-berlin.de}
\\\ \\
\inst{2} 
Fakult\"at f\"ur Mathematik und Informatik, \\
FernUniversit\"at in Hagen, Germany,
\\\ \\
}
\end{center}


\begin{abstract}
Two labeled point configurations $\{p_1,\ldots,p_n\}$ and $\{q_1,\ldots,q_n\}$ are of the same order type if, for every $i,j,k$, the triples $(p_i,p_j,p_k)$ and $(q_i,q_j,q_k)$ have the same orientation. 
In the 1980's, 
Goodman, Pollack and Sturmfels showed that
(i)~the number of order types on $n$ points is of order $4^{n + o(n)}$,
(ii)~all order types can be realized with double-exponential integer coordinates,
and that (iii)~certain order types indeed require double-exponential integer coordinates.
In 2018, 
Caraballo, D\'iaz-B\'a{\~n}ez, Fabila-Monroy, Hidalgo-Toscano, Lea{\~n}os, Montejano showed that at least  $n^{3n+o(n)}$ order types can be realized on an integer grid of polynomial size. 
In this article, we improve their result by showing that at least $n^{4n+o(n)}$ order types can be realized on an integer grid of polynomial size, which is essentially best possible.
\end{abstract}

\section{Introduction}

\begin{figure}[htb]
	\centering
	\includegraphics{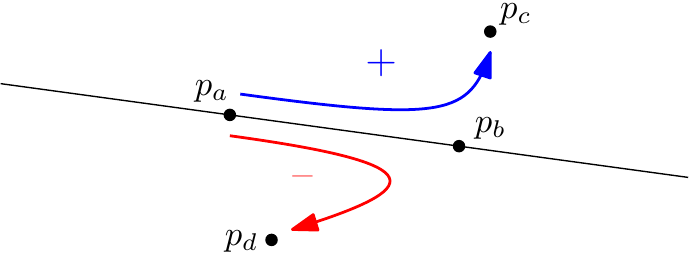}
	
	\caption{A chirotope with $\chi(a,b,c)=+$ and $\chi(a,b,d)=-$.}
	\label{fig:triple_orientations}  
\end{figure}

A set of $n$ labeled points $\{p_1,\ldots,p_n\}$ in the plane with $p_i=(x_i,y_i)$
induces a \emph{chirotope}, that is, a mapping $\chi \colon [n]^3 \to \{+,0,-\}$
which assigns an orientation $\chi(a,b,c) $ to each triple of points $(p_a,p_b,p_c)$ with
\[
\chi(a,b,c)  
= 
\sgn \det \begin{pmatrix}
1  & 1  & 1   \\
x_a& x_b& x_c \\
y_a& y_b& y_c\\
\end{pmatrix}.
\]
Geometrically this means  $\chi(a,b,c)$ is positive (negative) 
if the point $p_c$ lies to the left (right) of the directed line $\overrightarrow{p_ap_b}$ through $p_a$ directed towards $p_b$.
Figure~\ref{fig:triple_orientations} gives an illustration.
We say that two point sets are \emph{equivalent} 
if they induce the same chirotope 
and call the equivalence classes  \emph{order types}.
An order type in which
three or more points lie on a common line is called \emph{degenerate}.

Goodman and Pollack \cite{GoodmanPollack1983} (cf.\ \cite[Section~6.2]{Matousek2002_book}) showed
the number of order types on $n$ points is of order 
$\exp(4n\log n+O(n))=n^{4n+o(n)}$.
While the lower bound follows from a simple recursive construction 
of non-degenerate order types,
the proof of the upper bound uses the Milnor--Thom theorem \cite{Milnor1964,Thom1965} (cf.\ \cite{PetrovskiiOleinik1949,Warren1968}) 
- a powerful tool from real algebraic geometry.
The precise number of non-degenerate order types has been determined for up to 11 points
by Aichholzer, Aurenhammer, and Krasser  \cite{AichholzerAurenhammerKrasser2001,AichholzerKrasser2006} (cf.\ \cite{Knuth1992}).
For their investigations, they used computer-assistance to enumerate all ``abstract'' order types 
and  heuristics to either find a point set representation or to decide non-realizability.
Similar approaches have been taken by Fukuda, Miyata, and Moriyama \cite{FukudaMiyataMoriyama2013} (cf.\ \cite{FinschiFukuda2002})
to investigate order types with degeneracies for up to $8$ points. 
It is interesting to note that deciding realizability is an ETR-hard problem~\cite{Mnev1988}  and, since there are $\exp(\Theta(n^2))$ abstract order types (cf.\ \cite{DumitrescuMandal2019} and \cite{FelsnerValtr2011}), most of them are non-realizable.
For more details, we refer the interested reader
to the handbook article by Felsner and Goodman~\cite{FelsnerGoodman2016}.

Gr\"unbaum and Perles \cite[pp.~93--94]{Gruenbaum2005} (cf.\ \cite[pp.~355]{BjoenerLVWSZ1993}) showed that there exist degenerate order types 
that are only realizable with irrational coordinates.
Since we are mainly interested in order type representations with integer coordinates in this article,
we will restrict our attention in the following to the non-degenerate setting.

Goodman, Pollack, and Sturmfels \cite{GoodmanPollackSturmfels1989}
showed that all non-degenerate order types can be realized with double-exponential integer coordinates
and that certain order types indeed require double-exponential integer coordinates.
Moreover, from their construction one can also conclude that $n^{4n+o(n)}$ order types on $n$ points require integer coordinates of almost double-exponential size as outlined:
For a slowly growing function $f:\mathbb{N} \to \mathbb{R}$ with $f(n) \to \infty$ as $n \to \infty$ and $m=\lfloor n/f(n) \rfloor \ll n$,
we can combine each of the \mbox{$(n-m)^{4(n-m)+o(n-m)}=n^{4n+o(n)}$} 
order types of $n-m$ points with the $m$-point construction from \cite{GoodmanPollackSturmfels1989}, which requires integer coordinates of size
$\exp(\exp(\Omega(m)))$.
Another infinite family that requires integer coordinates of super-poly\-nomial size are the so-called \emph{Horton sets} \cite{BarbaDFMHT2017} (cf.\ \cite{Horton1983}),
which play a central role in the study of Erd\H{o}s--Szekeres--type \mbox{problems}.

In 2018,
Caraballo et al.\ \cite{CaraballoDBFMHTLM2018}
(cf.\ \cite{CaraballoDBFMHTLM2021})
showed that 
at least  $n^{3n+o(n)}$ non-degenerate order types can be realized on an integer grid of size $\Theta(n^{2.5}) \times \Theta(n^{2.5})$.
In this article, we follow a similar approach as in \cite{CaraballoDBFMHTLM2018} 
and improve their result by showing
that $n^{4n+o(n)}$ order types can be realized on a grid of  size \mbox{$\Theta(n^{4}) \times \Theta(n^{4})$}.

\begin{theorem}\label{thm:n_to_4n_many}
The number of non-degenerate order types which 
can be realized on an integer grid of size $(3n^4) \times (3n^4)$ 
is of order 
$n^{4n+o(n)}$.
\end{theorem}
An important consequence of Theorem~\ref{thm:n_to_4n_many} is that
a significant proportion of all $n$-point order types
can be stored as point sets with $\Theta( \log n)$ bits per point.
While the exponent in our $n^{4n+o(n)}$ bound is essentially best possible up to a lower-order error term, the question for the smallest constant $c$ remains open for which $n^{4n+o(n)}$ order types can be realized on a grid of size \mbox{$\Theta(n^c) \times \Theta(n^c)$}.

\paragraph{Related Work}
Besides the deterministic setting, 
also  integer grid representations of "random" order types have been intensively studied in the last years.
Fabila-Monroy and
Huemer  \cite{FMH2017} and
Devillers et al.\ \cite{DDGG2018} (cf.\ \cite{vdHMS2019}) independently showed that,
when a set $\{p_1,\ldots,p_n\}$ of real-valued points with $p_i=(x_i,y_i)$ are chosen uniformly and independently from the square $[0,n^{3+\epsilon}] \times [0,n^{3+\epsilon}]$, 
then the set $\{p_1',\ldots,p_n'\}$ with rounded (integer-valued) coordinates $p_i'=([x_i],[y_i])$ is of the same order type as the original set with high probability.

\goodbreak{}
\section{Proof of Theorem~\ref{thm:n_to_4n_many}}
\label{sec:proof_thm1}

Let $n$ be a sufficiently large positive integer.
It follows from Bertrand's postulate that
we can find a prime number $p$ 
satisfying $\frac{n}{2\lfloor \log n \rfloor} < p < \frac{n}{\lfloor \log n \rfloor}$.
As an auxiliary point set, we let 
\[
Q_p = \{ (x,y) \in \{1,\ldots,p \}^2 \colon y=x^2 \mod p \}.
\]
The point set $Q_p$ contains $p$ points from the $p \times p$ integer grid, and each two points have distinct $x$-coordinates. Moreover, $Q_p$ is non-degenerate 
because, by the Vandermonde determinant,
we have  
\[
\det
\begin{pmatrix}
1  & 1  & 1   \\
a& b& c \\
a^2& b^2& c^2\\ 
\end{pmatrix} = (b-a)(c-a)(c-b) \neq 0 \mod p,
\]
and hence $
\chi(a,b,c)  \neq 0$
for any pairwise distinct $a,b,c \in \{1,\ldots,p\}$.
In the following, we denote by $R(Q_p) = \{(y,x) \colon (x,y) \in Q_p\}$ the reflection of $Q_p$ with respect to the line $x=y$.

\begin{figure}[htb]
	\centering
	\includegraphics[width=0.9\textwidth]{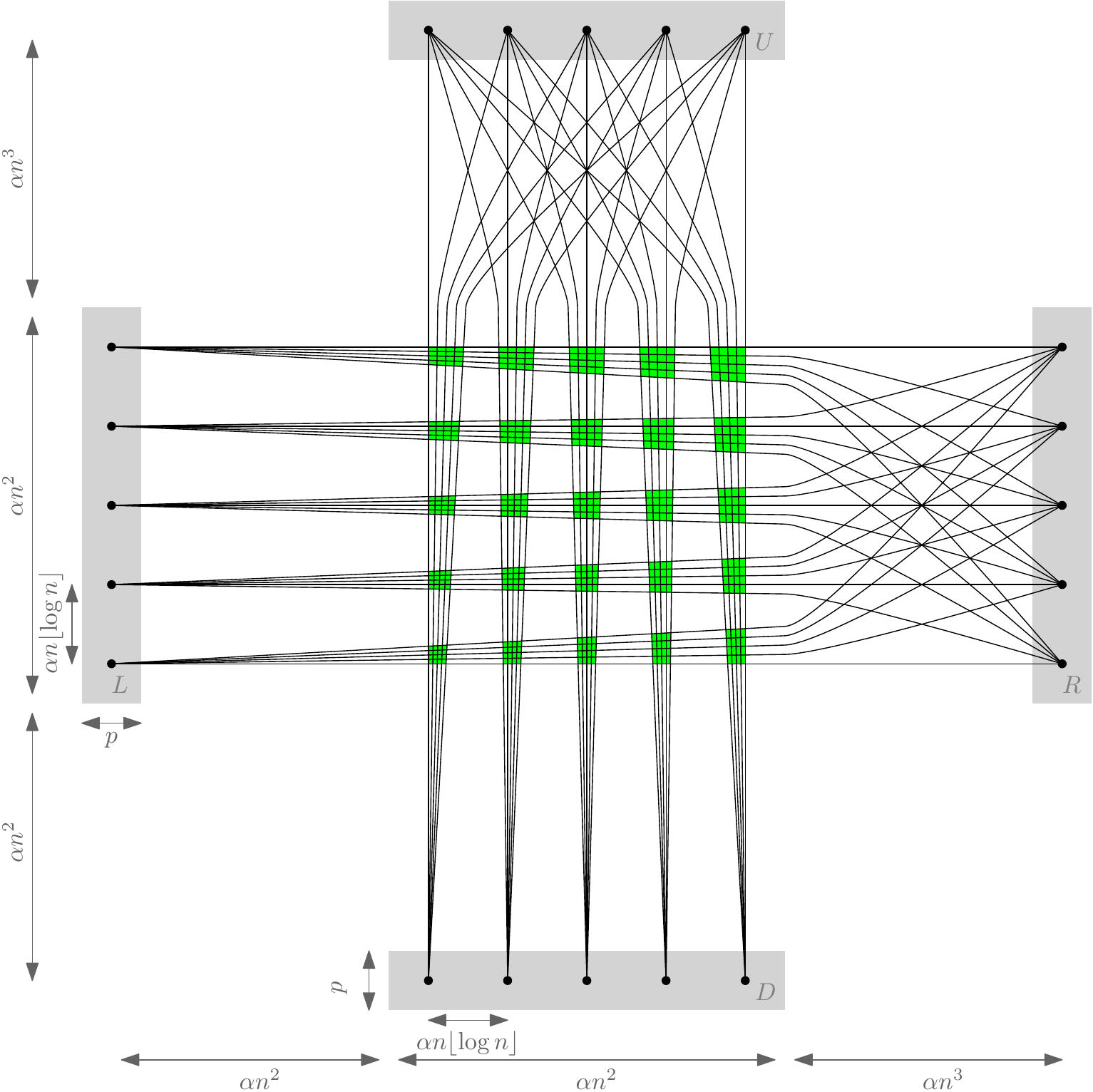}
	
	\caption{An illustration of the construction.
	The four copies $D,U,L,R$ of $Q_p$ are highlighted gray and
	the $(p^2-p)^2$ almost-square regions are highlighted green.}
	\label{fig:construction}  
\end{figure}

\medskip 
Let $\alpha = 2n$ and $m=\alpha \cdot (2n^2 + n^3)$.
For sufficiently large $n$, 
we have $2n^2 + n^3 \le 2n^3$ and 
$m + 2p \le 3 n^4$. 
Our goal is to construct $4^{n+o(n)}$ different
 $n$-point order types
on the integer grid 
\[
G = \{-p,\ldots,m+p\} \times \{-p,\ldots,m+p\}.
\]
We start with placing four scaled and translated copies of $Q_p$, which we denote by $D,U,L,R$,
as follows:
\begin{itemize}
\item
To obtain $D$, we scale $Q_p$ in $x$-direction by a factor $\alpha n \lfloor \log n \rfloor$ and translate by $(\alpha n^2,-p)$.
All points from $D$ 
have $x$-coordinates between $\alpha n^2$ and $2\alpha n^2$
and $y$-coordinates between $-p$ and $0$;
\item
To obtain $U$, we scale $Q_p$ in $x$-direction by a factor $\alpha n \lfloor \log n \rfloor$ and translate by $(\alpha n^2,m)$.
All points from $U$ 
have $x$-coordinates between $\alpha n^2$ and $2\alpha n^2$
and $y$-coordinates between $m$ and $m+p$;
\item
To obtain $L$, we scale $R(Q_p)$ in $y$-direction by a factor $\alpha n \lfloor \log n \rfloor$ and translate by $(-p,\alpha n^2)$.
All points from $L$ 
have $y$-coordinates between $\alpha n^2$ and $2\alpha n^2$
and $x$-coordinates between $-p$ and $0$;
\item
To obtain $R$, we scale $R(Q_p)$ in $y$-direction by a factor $\alpha n \lfloor \log n \rfloor$ and translate by $(m,\alpha n^2)$.
All points from $R$ 
have $y$-coordinates between $\alpha n^2$ and $2\alpha n^2$
and $x$-coordinates between $m$ and $m+p$.
\end{itemize}
Each pair of points $(l,r) \in L \times R$ 
spans an \emph{almost-horizontal} line-segment with absolute slope less than $\frac{1}{n}$.
Similarly, each pair of points from $D \times U$ spans an \emph{almost-vertical} line-segment with absolute reciprocal slope less than $\frac{1}{n}$.
As depicted in Figure~\ref{fig:construction}, 
these line-segments bound $(p^2-p)^2$ \emph{almost-square} regions.
Later, we will distribute the remaining $n-4p$ points 
among these almost-square regions in all possible way 
to obtain many different order types.

\medskip

For every pair of distinct points $d_1,d_2$ from $D$, 
the $x$-distance between them is at least $ \alpha n \lfloor \log n \rfloor$
and their $y$-distance is less than~$p$.
Hence, the absolute value of the slope of the line $\overline{d_1d_2}$ is less than $\frac{p}{\alpha n \lfloor \log n \rfloor}$.
Moreover, since $d_1$ and $d_2$ have non-positive $y$-coordinates and all points of $G$ are at $x$-distance  at most~$m$ from $d_1$ and~$d_2$, 
the line $\overline{d_1d_2}$ can only pass 
through points of $G$ with $y$-coordinate less than
$\frac{p \cdot m}{\alpha n \lfloor \log n \rfloor}  \le \alpha n^2$.
(Recall that $m=\alpha \cdot (2n^2 + n^3) \le 2 \alpha n^3$ and $p \le \frac{n}{\lfloor \log n \rfloor}$.)
We conclude that every point
from $U \cup L \cup R$ or from the almost-square regions
lies strictly above the line $\overline{d_1d_2}$.
Similar arguments apply to lines 
spanned by pairs of points from $U$, $L$, and~$R$, respectively.
Note that, in particular, our construction has the property that
for any point $q$ from an almost-square region, 
the point set $D \cup U \cup L \cup R \cup \{q\}$ 
is non-degenerate.

\goodbreak{}
\paragraph{Almost-square regions}

\begin{figure}[tb]
	\centering
	\includegraphics[page=2,width=0.9\textwidth]{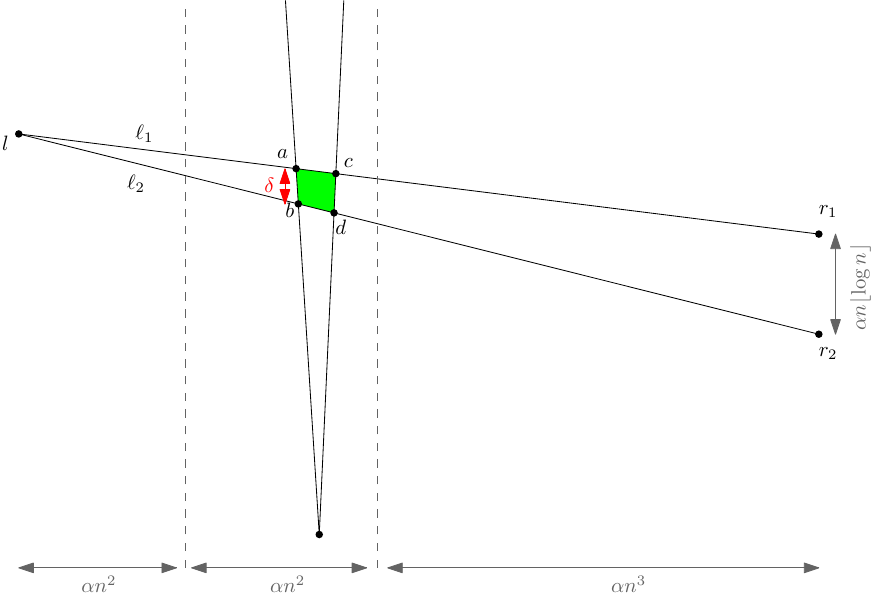}
	\caption{An almost-square region.}
	\label{fig:almostsquare}  
\end{figure}

Consider an almost-square region $A$ with 
top-left vertex~$a$, 
bottom-left vertex~$b$,
top-right vertex~$c$, 
and bottom-right vertex~$d$, 
as depicted in Figure~\ref{fig:almostsquare}.
By our construction, 
the two almost-horizontal line-segments $\ell_1,\ell_2$ bounding $A$ 
meet in a common end-point $l \in L$.
Let $r_1,r_2 \in R$ denote the other end-points of $\ell_1$ and $\ell_2$, respectively,
which have $y$-distance  $\alpha n \lfloor \log n \rfloor$.

The point $l$ has $x$-coordinate between $-p$ and $0$,
and  the $x$-coordinates of the two points $r_1$ and $r_2$ are between
$m$ and $m+p$. Since we have chosen $m=\alpha \cdot (2n^2 + n^3)$ and assumed $n$ to be sufficiently large, 
the $x$-distance between the points $l$ and $r_i$ (for $i=1,2$) 
is between $\alpha n^3$ and $2\alpha n^3$.
The point $a$ lies on $\ell_1$, $b$ lies on $\ell_2$, 
and the $x$-coordinates of both, $a$ and $b$, are between $\alpha n^2$ and $2\alpha n^2$.
Hence, we can bound the $y$-distance $\delta$ between $a$ and $b$ by
\[
\frac{1}{2}\alpha \lfloor \log n \rfloor
= 
\alpha n \lfloor \log n \rfloor \cdot \frac{\alpha n^2}{\alpha 2n^3} 
\le
\delta 
\le 
\alpha n \lfloor \log n \rfloor \cdot \frac{\alpha 2n^2}{\alpha n^3} 
=
2\alpha \lfloor \log n \rfloor
.
\]
Moreover, since $a$ and $b$ lie on an almost-vertical line (i.e., absolute reciprocal slope less than $\frac{1}{n}$), 
the $x$-distance  between $a$ and $b$ 
is less than $\frac{2 \alpha \lfloor \log n \rfloor}{n}$.
An analogous argument applies to the pairs $(a,c)$, $(b,d)$, and $(c,d)$,
and hence we can conclude that the almost-square region~$A$ contains at least 
\[
\left( \frac{1}{2}\alpha \lfloor \log n \rfloor - \frac{4 \alpha \lfloor \log n \rfloor}{n} \right)^2
\ge 
\frac{1}{5} \left( \alpha \lfloor \log n \rfloor  \right)^2
\]
 points from the integer grid~$G$, provided that $n$ is sufficiently large.

\goodbreak{}
\paragraph{Placing the remaining points}

We have already placed $p$ points in each of the four sets $D,U,L,R$.
For each of the remaining $n-4p$ points,
we can iteratively choose one of the $(p^2-p)^2$ almost-square regions 
and place it, unless our point set becomes degenerate.
To deal with these degeneracy-issue,
we denote an almost-square region $A$ \emph{alive}
if there is at least one point from $A$
which we can add to our current point configuration
while preserving non-degeneracy.
Otherwise we call $A$ \emph{dead}.

Having $k$ points placed ($4p \le k \le n-1$),
these $k$ points determine $\binom{k}{2}$ lines
which might \emph{kill} points from our integer grid 
and some almost-square regions become \emph{dead}.
That is, if we add another point that lies on one of these $\binom{k}{2}$ lines to our point configuration, 
we clearly have a degenerate order type.

To obtain a lower bound on the number of alive almost-square regions,
note that all almost-square regions lie in an 
$(\alpha n^2) \times (\alpha n^2)$ square
and that each of the $\binom{k}{2}$ lines kills 
at most $\alpha n^2$ grid points from almost-square regions.
Moreover, since each almost-square region contains at least $\frac{1}{5} \left( \alpha \lfloor \log n \rfloor  \right)^2$ grid points,
we conclude that the number of alive almost-square regions is at least
\[
(p^2-p)^2 - \binom{n}{2} \cdot \frac{\alpha n^2}{\frac{1}{5} \left( \alpha \lfloor \log n \rfloor  \right)^2} 
\ge 
\frac{1}{17}  \left(\frac{n}{\lfloor \log n \rfloor}\right )^4 
\]
for sufficiently large~$n$, 
since $\frac{n}{2\lfloor \log n \rfloor} \le p \le \frac{n}{\lfloor \log n \rfloor}$ and $\alpha = 2n$.

If we now place each of the remaining $n-4p$ points in an  almost-square region which is alive (one by one),
we have at least 
\[
\left( \frac{1}{17}  \left(\frac{n}{\lfloor \log n \rfloor}\right )^4 \right)^{n-4p} = n^{4n-O \left( n \frac{\log\log n}{\log n} \right)} 
\]
possibilities for doing so.
Each of these possibilities 
clearly gives us a different order type because, when we move a point $q$ 
from one almost-square region into another,
this point~$q$ moves over a line spanned by a pair 
$(l,r) \in L \times R$ or $(d,u) \in D \times U$,
and this affects $\chi(l,r,q)$ or $\chi(d,u,q)$, respectively.
This completes the proof of Theorem~\ref{thm:n_to_4n_many}.

{
	\small
	\bibliographystyle{alphaabbrv-url}
	\bibliography{bibliography}
}

\end{document}